\documentclass[10pt, aps, prd,
reprint, % for double column
notitlepage, nofootinbib, longbibliography, showpacs, letter, eqsecnum,
superscriptaddress,
]{revtex4-1}

\usepackage{pstricks}
\usepackage{amsmath,amssymb,bm}
\usepackage{pgfplots}
\usepackage{graphicx}
\usepackage{verbatim}
\usepackage[breaklinks=true,colorlinks=true,urlcolor=blue,linkcolor=red,citecolor=red]{hyperref}

\begin{document}

\title{Magnetostatic interaction energy between a point magnet and a ring magnet}

\author{Niranjan Warnakulasooriya}
\email{nwar@uttc.edu}
\affiliation{United Tribes Technical College, Bismarck, North Dakota 58504, USA}

\author{Dinuka H. Gallaba}
\email{dinuka4@gmail.com}
\affiliation{Department of Physics, Boston College, Chestnut Hill, Massachusetts 02467, USA}

\author{John Joseph Marchetta}
\email{jjmarchetta@gmail.com} 
\affiliation{Early Universe Cosmology and Strings (EUCOS) Group,
Center for Astrophysics, Space Physics and Engineering Research (CASPER),
Baylor University, Waco, Texas 76798, USA}
\affiliation{Department of Physics,
Baylor University, Waco, Texas 76798, USA}

\author{Duston Wetzel}
\email{duston.wetzel@siu.edu}
\affiliation{School of Physics and Applied Physics, Southern Illinois University--Carbondale,
Carbondale, Illinois 62901, USA}

\author{Prachi Parashar}
\email{Prachi.Parashar@jalc.edu} 
\affiliation{John A. Logan College, Carterville, Illinois 62918, USA}

\author{K. V. Shajesh}
\email{kvshajesh@gmail.com}
\affiliation{School of Physics and Applied Physics, Southern Illinois University--Carbondale,
Carbondale, Illinois 62901, USA}

\date{\today}

%--------------------------------------------
\begin{abstract}
We find an exact closed-form expression for the magnetostatic interaction
energy between a point magnet and a ring magnet in terms of complete
elliptic integrals. The exact expression for the energy exhibits an
equilibrium point close to the axis of symmetry of the ring magnet. Our
methodology will be useful in investigations concerning magnetic
levitation, and in the study of Casimir levitation.
\end{abstract}

\maketitle
%--------------------------------------------
%\tableofcontents
%--------------------------------------------
\section{Introduction}

Configurations with cylindrical symmetry often admit relatively simple
solutions on the axis of symmetry, even when the general solution
off the axis is given in terms of special functions or
has no exact solution.
A classic example is that of the magnetic field due to a circular wire
carrying a uniform current, where the expression for the magnetic
field on the axis is given in terms of rational functions and is
usually derived in an introductory level
physics course~\cite{Schwinger:1998cla}, while
the solution off the axis is given in terms of
complete elliptic integrals and is typically only introduced
in a graduate level course~\cite{Schwinger:1998cla}.

We show that the magnetostatic interaction energy between a point
magnet and a ring magnet also admits exact solutions in terms of
complete elliptic integrals when the point magnet is off the axis
of symmetry of the ring magnet and has a simple solution in terms of 
rational functions when the point dipole is on the axis of
the ring magnet. The interaction energy in general
exhibits an equilibrium point close to the axis of symmetry with a
saddle point instability. The expression for energy presented here
seems to have not been, to our surprise, reported before.
However, the corresponding expression for the magnetic field has been
discussed in the literature recently~\cite{Ravaud:2008rmp, Babic:2008rmp}.
The magnetic dipoles in their work are constructed by assuming the existence
of magnetic monopoles, which in the static case being considered
allows the use of the methodologies developed in electrostatics.
The methodology presented here is a useful academic exercise,
even though it presumes infinitely thin magnets.

We put forward two applications of the investigation presented here.
First is in the study of Casimir levitation.
The Casimir effect involves interactions
between materials with no net electric charge
and no permanent polarizations mediated by the electric and
magnetic fields induced from the quantum vacuum fluctuations.
Even though repulsion between anisotropically polarizable atoms were
well known~\cite{Axilrod:1943at,Muto:1943fc,Craig1969ma,Craig:1969pa,Babb2005ac},
perfectly conducting nanoparticles were not expected to show
repulsion from interactions with the quantum electromagnetic vacuum fluctuations.
Thus, it was a surprise when in Ref.~\cite{Levin:2010vo} it was shown that the interaction
between an anisotropically shaped conducting nanoparticle and a perfectly conducting 
metal sheet with a circular aperture could lead to repulsion.
Even though an analytic derivation of the result in Ref.~\cite{Levin:2010vo}
remains unsolved~\cite{Milton2011asa,2012:Miltonfpc,Shajesh2017ssa},
a partial understanding of the repulsion has been made plausible
by deriving analogous results in the non-retarded van der Waals regime~\cite{Eberlein2011hwp}
and in the retarded Casimir-Polder
regime~\cite{Milton2012esc,Shajesh:2011daa,Abrantes2018tcp,Marchetta:2020rps,Marchetta:2020dap}.
A drawback of all of the above investigations has been the
confinement of the nanoparticle to the axis of symmetry
in the configuration. Even though it is clear that the nanoparticle
is unstable in the transverse directions to the axis in the above
considerations, the limitation of being on the axis practically
does not allow any stability analysis. Before we embark on evaluating
the Casimir-Polder interaction energy between an anisotropically
polarizable nanoparticle and an anisotropically polarizable circular ring
without restricting the nanoparticle to being on the axis, we here explore
the analogous configuration of a permanent magnetic dipole moment
interacting with a circular ring with permanent polarization.
The methodology we use here can be immediately used
to study the corresponding Casimir interaction, which will be presented
elsewhere.

The second application is in the study of the magnetic levitation
of a Levitron$^\text{TM}$~\cite{Berry:1996sta}.
In particular, we would like to investigate
if the stability of the Levitron$^\text{TM}$ requires
the presence of gravity.
That is, can a spinning point magnet be
stabilized above a ring magnet in the absence of gravity?
The interaction energy presented here serves as the starting
point for this stability analysis.

In the next section we describe our configuration of a point magnet
and a ring magnet and
derive the expression for the interaction energy as an
integral over the azimuth angle.
In Section \ref{sec-com-ell-int} we give a brief description
of complete elliptic integrals. After introducing
complete elliptic integral
of the first kind $K(k)$ and second kind $E(k)$
we define elliptic integrals $\pi_3(k)$ and $\pi_5(k)$,
which is not the traditional approach. It should be possible
to express the elliptic integrals $\pi_3(k)$ and $\pi_5(k)$
in terms of the traditional elliptic integral of the third
kind. 
In Section \ref{sec-magEei} we derive the expression for
the interaction energy between a point magnet and a ring magnet
in terms of the elliptic integrals introduced in
Section \ref{sec-com-ell-int}.
In the final section we present our outlook concerning
the investigation of Casimir levitation.

%--------------------------------------------
\section{Magnetostatic energy}

Magnetostatics is governed by the Maxwell equations stating
that the magnetic field ${\bf B}({\bf r})$ is divergence free,
\begin{equation}
{\bm\nabla} \cdot {\bf B} =0,
\label{divFb}
\end{equation}
and that current densities ${\bf j}({\bf r})$ are sources
for the curl of the magnetic field,
\begin{equation}
{\bm\nabla} \times {\bf B} =\mu_0 {\bf j}.
\end{equation}
The conservation of charge in the static scenario requires the
current densities to be divergence free,
\begin{equation}
{\bm\nabla} \cdot {\bf j} =0.
\end{equation}
The constraint of a divergenceless magnetic field in Eq.\,(\ref{divFb})
allows the construction
\begin{equation}
{\bf B} = {\bm\nabla} \times {\bf A}
\end{equation}
in terms of the magnetic vector potential ${\bf A}({\bf r})$.
In conjunction with the Coulomb gauge,
\begin{equation}
{\bm\nabla} \cdot {\bf A} =0,
\end{equation}
this allows the solution for the vector potential
\begin{equation}
{\bf A}({\bf r}) =\frac{\mu_0}{4\pi} \int d^3r^\prime
\frac{{\bf j}({\bf r}^\prime)}{|{\bf r}-{\bf r}^\prime|}.
\end{equation}

The magnetic dipole moment of a given current density is
defined using the expression
\begin{equation}
{\bf m} =\frac{1}{2} \int d^3r^\prime
{\bf r}^\prime \times {\bf j}({\bf r}^\prime).
\end{equation}
For a circular current carrying loop of wire we have
$m=IA$, where $I$ is the current in the wire and $A$ is the
area of the circular loop. A point magnetic dipole is an
idealized construction with $I\to\infty$ and $A\to 0$,
keeping the product $m=IA$ fixed.
We shall be interested in the interaction between a point
magnetic dipole ${\bf m}_1$ and a ring magnet constructed
out of a uniform circular distribution of point dipoles ${\bf m}_2$.

The magnetic vector potential at position ${\bf r}$
due to a point magnetic dipole moment
${\bf m}_2$ placed at position ${\bf r}^\prime$ is
\begin{equation}
{\bf A}_2({\bf R}) = \frac{\mu_0}{4\pi}
\frac{{\bf m}_2 \times {\bf R}}{R^3},
\end{equation}
where
\begin{equation}
{\bf R} ={\bf r}-{\bf r}^\prime.
\label{def-Rrrp}
\end{equation}
The associated magnetic field due to the point magnet
is obtained using
\begin{equation}
{\bf B}_2 = {\bm\nabla} \times {\bf A}_2
\end{equation}
and leads to the expression
\begin{equation}
{\bf B}_2({\bf R}) = \frac{\mu_0}{4\pi}
\frac{\big[ 3 \hat{\bf R} \hat{\bf R}
-{\bf 1} \big] \cdot {\bf m}_2}{R^3},
\qquad {\bf r} \neq {\bf r}^\prime,
\label{B-di-e}
\end{equation}
where $\hat{\bf R} ={\bf R}/R$.
This expression for the magnetic field
in Eq.\,(\ref{B-di-e}) is missing a term
$\mu_0 {\bf m}_2 \delta^{(3)}({\bf r}-{\bf r}^\prime)$
which contributes only at ${\bf r} = {\bf r}^\prime$
and is necessary to satisfy the constraint
\begin{equation}
{\bm\nabla} \cdot {\bf B}_2=0.
\end{equation}
The magnetostatic interaction energy
between another point magnetic dipole ${\bf m}_1$
and the dipole ${\bf m}_2$ is given by
\begin{equation}
U({\bf r}) = -{\bf m}_1 \cdot {\bf B}_2({\bf r}),
\end{equation}
where ${\bf r}$ now is the position of the
point magnet ${\bf m}_1$.

%-------
\begin{figure}
\includegraphics{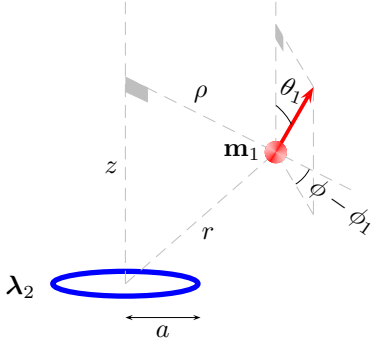}
\caption{
A point magnet of magnetic dipole moment ${\bf m}_1=m_1\hat{\bf n}$
at height $z$ above a ring magnet of radius $a$ with
uniform magnetic dipole moment per unit length ${\bm\lambda}_2$.
The point magnet is a distance $\rho$ away from the axis of symmetry
of the ring magnet.
The dipole moment subtends an angle $\theta_1$ with respect
to the axis of symmetry, that is,
$\hat{\bf n} \cdot\hat{\bf z} = \cos\theta_1$.
}
\label{atom-above-dielectric-ring-spin-fig}
\end{figure}%
%-------

A ring magnet is described by its magnetic moment per unit length
\begin{equation}
{\bm\lambda}_2 = \frac{d{\bf m}_2}{ad\phi},
\label{ring-dpl-def}
\end{equation}
where $a$ is the radius of the ring and $ad\phi$ is the 
differential arc length. Let us choose the magnetic moment
of the ring to be uniform and along the axis of symmetry
of the ring, say $\hat{\bf z}$, such that
\begin{equation}
{\bm\lambda}_2 = \lambda_2 \hat{\bf z}.
\end{equation}
We further choose the ring to be in the $z=0$ plane
centered at the origin.
Refer Fig.~\ref{atom-above-dielectric-ring-spin-fig}.
Let us keep the orientation of the point magnet arbitrary
relative to the ring magnet and describe it as
\begin{equation}
{\bf m}_1 = m_1 \hat{\bf n},
\end{equation}
where
\begin{equation}
\hat{\bf n} = \sin\theta_1\cos\phi_1 \hat{\bf x}
+\sin\theta_1\sin\phi_1 \hat{\bf y} + \cos \theta_1 \hat{\bf z},
\label{def-nuv}
\end{equation}
such that
\begin{equation}
\hat{\bf n} \cdot \hat{\bf z} =\cos\theta_1
\label{ndz=ct1}
\end{equation}
with its position
\begin{equation}
{\bf r} =  \rho\cos\phi \hat{\bf x}
+ \rho\sin\phi \hat{\bf y}
+ z \hat{\bf z}.
\label{def-rpv}
\end{equation}
Note that
\begin{equation}
\hat{\bf n} \cdot {\bf r }
 =\rho\sin\theta_1\cos(\phi-\phi_1),
\end{equation}
which illustrates that the vectors ${\bf m}_1$ and
${\bm\lambda}_2$ representing the orientation of the
dipoles and ${\bf r}$ are not in the same plane.

Differential contribution to the interaction energy from the
interaction between the point magnet and a differential 
section of the ring magnet is given by
\begin{equation}
dU = -{\bf m}_1 \cdot d{\bf B}_2,
\end{equation}
where using Eq.\,(\ref{B-di-e})
\begin{equation}
d{\bf B}_2({\bf R}) = \frac{\mu_0}{4\pi}
\frac{\big[ 3 \hat{\bf R} \hat{\bf R}
-{\bf 1} \big] \cdot d{\bf m}_2}{R^3}
\end{equation}
with ${\bf r}^\prime$ now constrained to be on the ring by
$z^\prime=0$ and $|{\bf r}^\prime|=a$ such that
\begin{equation}
{\bf r}^\prime =  a\cos\phi^\prime \hat{\bf x}
+ a\sin\phi^\prime \hat{\bf y}
+ 0 \hat{\bf z}.
\label{def-rppv}
\end{equation}
Using Eq.\,(\ref{ring-dpl-def})
the differential interaction energy takes the form
\begin{equation}
dU = \frac{\mu_0}{4\pi}
\frac{{\bf m}_1 \cdot \big[ 
{\bf 1} -3 \hat{\bf R} \hat{\bf R} \big] 
\cdot {\bm\lambda}_2}{R^3} ad\phi^\prime
\end{equation}
from which the total interaction energy can be
calculated by integrating over angle $\phi^\prime$
and is given by
\begin{equation}
U = \frac{\mu_0}{4\pi} m_1\lambda_2
\int_0^{2\pi} ad\phi^\prime
\left[ \frac{(\hat{\bf n} \cdot \hat{\bf z})}{R^3}
-\frac{(\hat{\bf n} \cdot {\bf R})({\bf R} \cdot \hat{\bf z}}{R^5}
\right],
\end{equation}
where
\begin{equation}
R = \sqrt{z^2 + a^2 +\rho^2 -2a\rho \cos(\phi^\prime-\phi)}.
\label{def-Rmag}
\end{equation}
We have $(\hat{\bf n} \cdot \hat{\bf z}$)
using Eq.\,(\ref{ndz=ct1}),
\begin{equation}
{\bf R} \cdot \hat{\bf z}=z,
\label{Rdz=z}
\end{equation}
and
\begin{equation}
\hat{\bf n} \cdot {\bf R}
=\rho\sin\theta_1\cos(\phi-\phi_1)
-a\sin\theta_1\cos(\phi^\prime-\phi_1) +z\cos\theta_1.
\label{ndR=ex}
\end{equation}
Using these expressions the magnetostatic
interaction energy between the point magnet
and the ring magnet is given by
\begin{eqnarray}
&& U(z,\rho,\phi-\phi_1,\theta_1)
= \frac{\mu_0}{4\pi} \frac{m_1(2\pi\lambda_2)}{a^2}
\int_0^{2\pi} \frac{d\phi^\prime}{2\pi}
\nonumber \\ && \times 
\Bigg[ \frac{a^3 \cos\theta_1}{R^3}
- \frac{3a^3z^2 \cos\theta_1}{R^5}
-\frac{3a^3 z\rho\sin\theta_1 \cos(\phi-\phi_1)}{R^5}
 \nonumber \\
&& \hspace{4mm}
+\frac{3a^4 z\sin\theta_1 \cos(\phi^\prime-\phi_1)}{R^5}
\Bigg]. \hspace{8mm}
\label{ie-ex34}
\end{eqnarray}
In the special circumstance when the point magnet is
positioned on the axis of the ring we have $\rho=0$.
This allows the integrals on $\phi^\prime$ in Eq.~(\ref{ie-ex34})
to be completed and yields an exact expression for
the interaction energy for this scenario as
\begin{equation}
U(z,0,\phi-\phi_1,\theta_1)
= \frac{\mu_0}{4\pi} \frac{m_1(2\pi\lambda_2)}{a^2}
\frac{a^3(a^2-2z^2)}{(a^2 +z^2)^\frac{5}{2}} \cos\theta_1,
\label{ier=034}
\end{equation}
which has an extremum at
\begin{equation}
z = h = \pm a\sqrt{\frac{3}{2}}.
\end{equation} 
When the point magnet is positioned at this extremum point
$z=h$ on the axis we have
\begin{equation}
U(h,0,\phi-\phi_1,\theta_1)
= -\frac{\mu_0}{4\pi} \frac{m_1(2\pi\lambda_2)}{a^2}
\frac{8}{25} \sqrt{\frac{2}{5}} \cos\theta_1.
\end{equation}
In general, for $\rho\neq 0$, the integrals on $\phi^\prime$
can not be completed in terms of elementary functions.
However, they can be expressed in terms of complete elliptic
integrals. 
In the following section, we shall evaluate the exact and
approximate form for the elliptic integrals required to
express Eq.\,(\ref{ie-ex34}) for $\rho\neq 0$ off the axis.

%-----------------------------------------------------
\section{Complete elliptic integrals}
\label{sec-com-ell-int}

\begin{figure}
\includegraphics{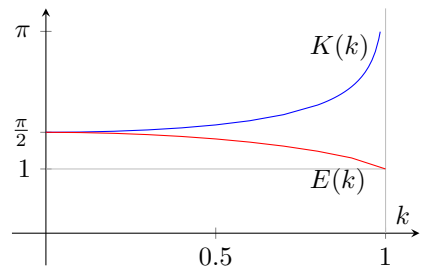}
\caption{Complete elliptic integrals of the first
kind $K(k)$ and of the second kind $E(k)$, plotted
as a function of $k$. Both the functions evaluate to
$\pi/2$ for $k=0$. For $k\to 1$ the elliptic integral
of the second kind approaches 1 and the elliptic
integral of the first kind grows logarithmically.}
\label{comEI-fig}
\end{figure}

Complete elliptic integrals of the first and second kind can be defined using
the integral representations~\cite{NIST:DLMF,NIST:2010fm}
\begin{subequations}
\begin{eqnarray}
K(k) &=& \int_0^\frac{\pi}{2} d\psi \frac{1}{\sqrt{1-k^2\sin^2\psi}}, \\
E(k) &=& \int_0^\frac{\pi}{2} d\psi \sqrt{1-k^2\sin^2\psi},
\end{eqnarray}%
\label{ComEllInt12}%
\end{subequations}%
respectively. We will be interested in the domain $0\leq k<1$.
These integrals can not be completed and expressed
in terms of elementary functions. However, for special values
they can be evaluated easily. For example, we can verify that 
\begin{subequations}
\begin{eqnarray}
K(0) &=& \frac{\pi}{2}, \\
E(0) &=& \frac{\pi}{2}.
\end{eqnarray}
\end{subequations}
Further, we can verify that
\begin{equation}
E(1) = 1.
\end{equation}
Note that
\begin{equation}
K(1) = \int_0^\frac{\pi}{2} \frac{d\psi}{\cos\psi}
\end{equation}
is divergent. To see the nature of this divergence we can introduce a cutoff
parameter $\delta >0$ and write
\begin{equation}
K(1) = \lim_{\delta\to 0}
 \int_0^{\frac{\pi}{2}-\delta} \frac{d\psi}{\cos\psi},
\end{equation}
which when evaluated using the identity
$d( \sec\psi +\tan\psi)= \sec\psi ( \sec\psi +\tan\psi)d\psi$
yields 
\begin{equation}
K(1) \sim \ln 2 - \ln\delta -\frac{\delta^2}{12} + {\cal O}(\delta)^4
\end{equation}
and reveals that $K(1)$ has a logarithmic divergence.
The plots of $K(k)$ and $E(k)$ as functions of $k$ for $0\leq k<1$
are shown in Fig.~\ref{comEI-fig}.
The complete elliptic integrals in Eqs.\,(\ref{ComEllInt12})
have the power series expansions
\begin{subequations}
\begin{eqnarray}
K(k) &=& \frac{\pi}{2} \sum_{n=0}^\infty
\left[ \frac{(2n)!}{2^{2n} (n!)^2} \right]^2 k^{2n} \\
&=& \frac{\pi}{2} \left[
1 + \frac{1}{4} k^2 +\frac{9}{64} k^4 + \ldots \right], \\
E(k) &=& \frac{\pi}{2} \sum_{n=0}^\infty
%\left[ \frac{(2n)!}{2^{2n} (n!)^2} \right]^2 k^{2n}
\left[ \frac{(2n)!}{2^{2n} (n!)^2} \right]^2 \frac{k^{2n}}{(1-2n)} \\
&=& \frac{\pi}{2} \left[
1 - \frac{1}{4} k^2 -\frac{3}{64} k^4 - \ldots \right].
\end{eqnarray}
\end{subequations}
The leading order contribution in the power series expansions
are from $K(0)$ and $E(0)$.
The next-to-leading order contributions in the above
series expansions are evaluated by expanding the radical in
Eqs.(\ref{ComEllInt12}) as a series using
\begin{subequations}
\begin{eqnarray}
\frac{1}{\sqrt{1-x}} &=& 1 +\frac{1}{2} x +\dots, \\
\sqrt{1-x} &=& 1 -\frac{1}{2} x +\dots.
\end{eqnarray}
\end{subequations}
Either the integral representations or the series expansions
are sufficient to investigate the properties of
the complete elliptic integrals. Here we shall primarily
use the integral representations, and depend on the series
expansions occasionally.

%-------
%-------

To get some insight for complete elliptic integrals we
mention three physical situations where one encounters
these functions. Firstly,
if we had sought to evaluate the perimeter of an ellipse
during our exposure to geometry, we would have
encountered the complete elliptic integral of the second kind.
The perimeter $C$ of an ellipse, described by the equation
\begin{equation}
\frac{x^2}{a^2} +\frac{y^2}{b^2} =1
\end{equation}
and characterized by the eccentricity
\begin{equation}
e = \sqrt{1 - \frac{b^2}{a^2}}
\end{equation}
in terms of the semi-major axis $a$ and semi-minor axis $b$,
is given in terms of complete elliptic integral of the second kind as 
\begin{equation}
C = 4a E(e).
\end{equation}
A circle is an ellipse of zero eccentricity ($a=b$)
and has the circumference
\begin{equation}
C \to 4a E(0) =2\pi a
\end{equation}
using $E(0)=\pi/2$.
Secondly, the period of oscillations $T$ of the simple pendulum
as a function of the amplitude of oscillations $\phi_0$ is
given in terms of the complete elliptic integral of the first kind as 
\begin{equation}
T = 2\pi\sqrt{\frac{l}{g}}
 \,\frac{2}{\pi} K\left(\sin\frac{\phi_0}{2}\right).
\label{T=Kfun}
\end{equation}
For small amplitudes ($\phi_0\ll 1$) this reproduces the  
classic result
\begin{equation}
T \to 2\pi\sqrt{\frac{l}{g}}
 \,\frac{2}{\pi} K\left(0 \right) = 2\pi\sqrt{\frac{l}{g}}
\end{equation}
using $K(0)=\pi/2$.
Thirdly, one encounters elliptic integrals while
finding the magnetic field due to a circular wire carrying a
steady current, at points away from the axis of symmetry of the
circular wire~\cite{Schwinger:1998cla}.

Derivatives of the elliptic integrals with respect to their
arguments are calculated by evaluating the derivatives of the
corresponding integrands and then rewriting the resultant integrals
in terms of elliptic integrals. This process is simplified by
introducing new elliptic integrals. 
The derivative of the complete elliptic
integral of the second kind leads to the integral 
\begin{equation}
\frac{dE}{dk} = -k\int_0^\frac{\pi}{2} d\psi
\frac{\sin^2\psi}{\sqrt{1-k^2\sin^2\psi}},
\label{dE=s21}
\end{equation} 
which can be rewritten in the form
\begin{equation}
\frac{dE}{dk} = \frac{1}{k}\int_0^\frac{\pi}{2} d\psi
\frac{\big[-1+1-k^2\sin^2\psi\big]}{\sqrt{1-k^2\sin^2\psi}}
\end{equation}
to recognize the identity
\begin{equation}
\frac{dE}{dk} = -\frac{K(k)}{k} +\frac{E(k)}{k}.
\label{dE=idty}
\end{equation}
Following the same steps for $K(k)$ yields
\begin{equation}
\frac{dK}{dk} = \frac{\pi_3(k)}{k} -\frac{K(k)}{k},
\label{dK=idty}
\end{equation}
where we introduced a new elliptic integral
\begin{equation}
\pi_3(k) = \int_0^\frac{\pi}{2} d\psi
\frac{1}{(1-k^2\sin^2\psi)^\frac{3}{2}}.
\label{pi3-def}
\end{equation}

The new elliptic integral $\pi_3(k)$ can be written
in terms of $K(k)$ and $E(k)$. To obtain this result,
we rewrite the integral in Eq.\,(\ref{dE=s21}) in the form
\begin{equation}
\frac{dE}{dk} = k\int_0^\frac{\pi}{2} d\psi
\frac{\sin\psi}{\sqrt{1-k^2\sin^2\psi}}
\frac{d}{d\psi} \cos\psi
\label{s2d=-sdc}
\end{equation}
and integrate by parts to write
\begin{eqnarray}
\frac{dE}{dk} &=& k\int_0^\frac{\pi}{2} d\psi
\frac{d}{d\psi} \left[ 
\frac{\sin\psi\cos\psi}{\sqrt{1-k^2\sin^2\psi}} \right]
\nonumber \\
&&-k\int_0^\frac{\pi}{2} d\psi \cos\psi \frac{d}{d\psi}
\left[ \frac{\sin\psi}{\sqrt{1-k^2\sin^2\psi}} \right].
\hspace{3mm}
\end{eqnarray}
The first integrand is a total derivative and thus
contributes only at the boundary, and yields zero in
this case at both ends. The second integral, after
evaluating the derivative in the integrand, takes the form
\begin{equation}
\frac{dE}{dk} = \int_0^\frac{\pi}{2} d\psi
\frac{-k\cos^2\psi}{(1-k^2\sin^2\psi)^\frac{3}{2}}.
\end{equation}
Rewriting the numerator of the integrand as
\begin{equation}
-k\cos^2\phi = \frac{(1-k^2)}{k}
-\frac{(1-k^2\sin^2\psi)}{k}
\end{equation}
allows us to recognize the integrals as
\begin{equation}
\frac{dE}{dk} = \pi_3(k) \frac{(1-k^2)}{k}
-\frac{K(k)}{k}.
\label{dE=idty2}
\end{equation}
Thus, we have derived two separate expressions for
$dE/dk$ in Eqs.\,(\ref{dE=idty}) and (\ref{dE=idty2}).
Equating the right hand sides of these equations allows
us to find an identity for $\pi_3(k)$ in terms of $E(k)$,
\begin{equation}
\pi_3(k) = \frac{E(k)}{(1-k^2)}.
\label{pi3-ito-E}
\end{equation}
Using the power series expansion for $E(k)$ together with
the power series expansion of $1/(1-k^2)$ we obtain the
power series expansion for $\pi_3(k)$ as
\begin{equation}
\pi_3(k) = \frac{\pi}{2} \left[
1 +\frac{3}{4} k^2 +\frac{45}{64} k^4 +\ldots \right].
\label{pi3s-3t}
\end{equation}

When we follow the steps leading to Eq.\,(\ref{dE=idty})
for $\pi_3(k)$ we obtain
\begin{equation}
\frac{d\pi_3}{dk} = \frac{3}{k}
\Big[ \pi_5(k) -\pi_3(k) \Big],
\end{equation}
where
\begin{equation}
\pi_5(k) = \int_0^\frac{\pi}{2} d\psi
\frac{1}{(1-k^2\sin^2\psi)^\frac{5}{2}}.
\label{pi5-def}
\end{equation}
Starting from the definition of $K(k)$ we have the derivative
\begin{equation}
\frac{dK}{dk} = k\int_0^\frac{\pi}{2} d\psi
\frac{\sin^2\psi}{(1-k^2\sin^2\psi)^\frac{3}{2}}.
\label{dK=s21}
\end{equation}
Using the identity $\sin^2\psi d\psi =-\sin\psi d\cos\psi$,
like earlier in Eq.\,(\ref{s2d=-sdc}), we integrate by parts
to obtain
\begin{equation}
\frac{dK}{dk} = \int_0^\frac{\pi}{2} d\psi
\frac{k\cos^2\psi(1+2k^2\sin^2\psi)}{(1-k^2\sin^2\psi)^\frac{5}{2}}.
\end{equation}
Again, rewriting the numerator as
\begin{eqnarray}
&& \cos^2\psi(1+2k^2\sin^2\psi)
=-\frac{3(1-k^2)}{k^2}
\hspace{28mm} \nonumber \\ && \hspace{3mm}
+\frac{(5-2k^2)}{k^2} (1-k^2\sin^2\psi)
-\frac{2}{k^2} (1-k^2\sin^2\psi)^2
\hspace{5mm}
\end{eqnarray}
leads to the identity
\begin{equation}
\frac{dK}{dk} = -\frac{3(1-k^2)}{k^2} \pi_5(k)
+\frac{(5-2k^2)}{k^2} \pi_3(k) -\frac{2}{k^2} K(k). 
\label{dK=s2m}
\end{equation}
Using Eqs.\,(\ref{dK=idty}) and (\ref{dK=s2m}) we have
\begin{equation}
\pi_5(k) = \frac{2(2-k^2)}{3(1-k^2)} \pi_3(k)
-\frac{K(k)}{3(1-k^2)}.
\end{equation}
We can further replace $\pi_3(k)$ 
Eq.\,(\ref{pi3-ito-E}) to write
\begin{equation}
\pi_5(k) = \frac{2(2-k^2)}{3(1-k^2)^2} E(k)
-\frac{K(k)}{3(1-k^2)}.
\end{equation}
The power series expansion for $\pi_5(k)$ yields
\begin{equation}
\pi_5(k) = \frac{\pi}{2} \left[
1 +\frac{5}{4} k^2 +\frac{105}{64} k^4 +\ldots \right].
\label{pi5s-3t}
\end{equation}
For the present discussion it is also handy
to have the series expansion
\begin{eqnarray}
&&\left( \pi_5(k) -\frac{2}{k^2}
\Big\{ \pi_5(k) -\pi_3(k) \Big\} \right)
\hspace{30mm} \nonumber \\ && \hspace{20mm}
= \frac{\pi}{2} \Big[0 
-\frac{5}{8} k^2 -\frac{35}{32} k^4 +\ldots \Big].
\label{pi35s-3t}
\end{eqnarray}

%-----------------------------------------------------
\section{Magnetostatic energy in terms of complete
elliptic integrals}
\label{sec-magEei}

To express the magnetostatic interaction energy in
Eq.\,(\ref{ie-ex34}) in terms of elliptic integrals,
we start by substituting $\phi^{\prime\prime}=\phi^\prime-\phi$,
which takes the limit of integrations from $-\phi$ to
$2\pi -\phi$. Since the integration is a sum, it does not care for the
order as long as it completes a period. Thus, we can
switch the limits of integration to go from $-\pi$ to $+\pi$.
This leads to
\begin{eqnarray}
U &=& \frac{\mu_0}{4\pi} \frac{m_1(2\pi\lambda_2)}{a^2}
\int_{-\pi}^{\pi} \frac{d\phi^{\prime\prime}}{2\pi}
\Bigg[ \frac{a^3 \cos\theta_1}{R^3}
- \frac{3a^3z^2 \cos\theta_1}{R^5}
\nonumber \\ &&
-\frac{3a^3 z\rho\sin\theta_1 \cos(\phi-\phi_1)}{R^5}
\nonumber \\ &&
+\frac{3a^4 z\sin\theta_1 \cos(\phi^{\prime\prime}+\phi-\phi_1)}{R^5}
\Bigg],
\end{eqnarray}
where, now, $R^2=z^2+a^2+\rho^2-2a\rho\cos\phi^{\prime\prime}$.
The integral associated with the fourth term evaluates partly to zero,
after using $ \cos(\phi^{\prime\prime}+\phi-\phi_1)
=\cos\phi^{\prime\prime}\cos(\phi-\phi_1)
-\sin\phi^{\prime\prime}\sin(\phi-\phi_1)$,
because the integrand containing $\sin\phi^{\prime\prime}$
is odd, and the rest being even are twice the value when integrating from
0 to $\pi$. Thus,
\begin{eqnarray}
U &=& \frac{\mu_0}{4\pi} \frac{m_1(2\pi\lambda_2)}{a^2}
\int_0^\pi \frac{d\phi^{\prime\prime}}{2\pi}
\Bigg[ \frac{2a^3 \cos\theta_1}{R^3}
- \frac{6a^3z^2 \cos\theta_1}{R^5}
\nonumber \\ && 
-\frac{6a^3 z\sin\theta_1 \cos(\phi-\phi_1)(\rho -a\cos\phi^{\prime\prime})}{R^5}
\Bigg].
\end{eqnarray}
To prepare the denominator for the elliptic integrals
we substitute $\phi^{\prime\prime}=\pi-\phi^\prime$,
which amounts to integrating in the reverse order. This
amounts to replacing 
$\cos\phi^\prime \to\cos(\pi-\phi^\prime) =
-\cos\phi^\prime$. That is,
\begin{eqnarray}
U &=& \frac{\mu_0}{4\pi} \frac{m_1(2\pi\lambda_2)}{a^2}
\int_0^\pi \frac{d\phi^\prime}{2\pi}
\Bigg[ \frac{2a^3 \cos\theta_1}
{(a^2+z^2+\rho^2+2 a\rho \cos\phi^\prime)^\frac{3}{2}} \nonumber \\ &&
-\frac{6a^3z^2 \cos\theta_1}
{(a^2+z^2+\rho^2+2 a\rho \cos\phi^\prime)^\frac{5}{2}}
\nonumber \\
&& - \frac{6a^3 z\sin\theta_1 \cos(\phi-\phi_1)
(\rho+a\cos\phi^\prime)} {(a^2+z^2+\rho^2+2 a\rho \cos\phi^\prime)^\frac{5}{2}} \Bigg].
\end{eqnarray}
Using the trigonometric identity
$\cos\phi^\prime =1-2\sin^2(\phi^\prime/2)$
and substituting $\phi^\prime/2 \to \phi^\prime$
afterwards, we obtain
\begin{eqnarray}
U &=& \frac{\mu_0}{4\pi} \frac{m_1(2\pi\lambda_2)}{a^2}
\sqrt{\frac{a}{\rho}}
\int_0^\frac{\pi}{2} \frac{d\psi}{2\pi} \Bigg[
\frac{ak^3\cos\theta_1}{2\rho(1-k^2\sin^2\psi)^\frac{3}{2}}
\nonumber \\ &&
-\frac{3z^2k^5\cos\theta_1}{8\rho^2(1-k^2\sin^2\psi)^\frac{5}{2}}
%\nonumber \\ && 
- \frac{3zak^5\sin\theta_1 \cos(\phi-\phi_1)}
{8\rho^2(1-k^2\sin^2\psi)^\frac{5}{2}}
\nonumber \\ && \times
\left\{ \frac{\rho}{a} +(1-2\sin^2\psi) \right\}
\Bigg]. 
\end{eqnarray}
We can recognize the elliptic integrals
$\pi_3(k)$ and $\pi_5(k)$ introduced in
Eqs.\,(\ref{pi3-def}) and (\ref{pi5-def}), respectively,
in the first two integrals and in the first term of the 
third integral. 
The elliptic integrals
here are written in terms of the argument $k$ defined using
\begin{equation}
k^2 = \frac{4a\rho}{z^2+(a+\rho)^2}.
\label{def-k2}
\end{equation}
The second term in the third integral can be expressed in
terms of elliptic integrals as
\begin{eqnarray}
&& \int_0^\frac{\pi}{2} d\psi
\frac{(1-2\sin^2\psi)}{(1-k^2\sin^2\psi)^\frac{5}{2}}
\hspace{30mm} \nonumber \\ && \hspace{10mm} 
= \left( \pi_5(k) -\frac{2}{k^2}
\Big\{ \pi_5(k) -\pi_3(k) \Big\} \right).
\end{eqnarray}
Then, in terms of elliptic integrals,
%$\pi_3(k)$ and $\pi_5(k)$
%introduced earlier  in Eqs.\,(\ref{pi3-def}) and (\ref{pi5-def}),
we obtain an exact analytic expression for the magnetostatic 
interaction energy between the point dipole and the ring magnet as
\begin{widetext}
\begin{eqnarray}
U(z,\rho,\phi-\phi_1,\theta_1)
&=& \frac{\mu_0}{4\pi} \frac{m_1(2\pi\lambda_2)}{a^2}
\sqrt{\frac{a}{\rho}} \frac{1}{4} 
\frac{2}{\pi} \Bigg[ 
\frac{1}{2} \frac{a}{\rho} \cos\theta_1 k^3\pi_3(k)
-\frac{3}{8} \left( \frac{z^2}{\rho^2} \cos\theta_1
+\frac{z}{\rho} \sin\theta_1 \cos(\phi-\phi_1)
\right) k^5\pi_5(k)
\nonumber \\ && \hspace{34mm}
-\frac{3}{8} \frac{a}{\rho} \frac{z}{\rho} \sin\theta_1 \cos(\phi-\phi_1)
k^5 \left( \pi_5(k) -\frac{2}{k^2}
\Big\{ \pi_5(k) -\pi_3(k) \Big\} \right) \Bigg].
\label{Eie-56}
\end{eqnarray}
\end{widetext}
The expression for the interaction energy in Eq.\,(\ref{Eie-56})
is valid for arbitrary position and orientation
of the point magnet. We shall proceed to list some
special cases of positions and orientations, which
are expected to give insight into the structure of the
interaction energy.

%-------
\begin{figure}
\includegraphics[width=8.5cm]{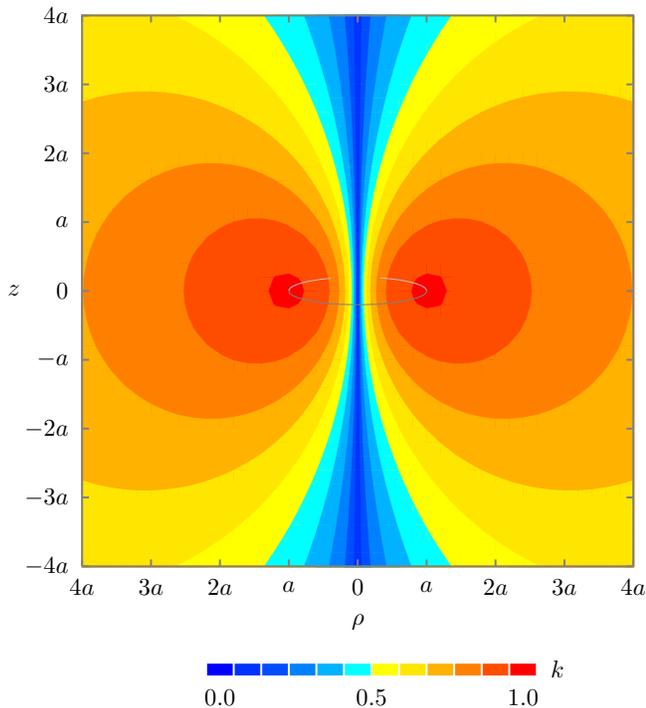}
\caption{Contour plot of the parameter
$k$ defined in Eq.\,(\ref{def-k2}) as a function of $\rho$ and $z$.
In the figure, $k=0$ corresponds to the $z$ axis where $\rho=0$,
and $k=1$ corresponds to the ring described by $\rho=a$.
The region corresponding to $k\ll 1$ consists of points very
close to the $z$ axis. }
\label{ksquare-fig}
\end{figure}%
%-------

In the special case when the point
magnet is positioned on the axis of symmetry of the ring
magnet we have $\rho=0$, which sets $k=0$.
We keep the orientation of the point magnet arbitrary.
The parameter $0\leq k<1$ spans the complete region
around the ring magnet. $k=1$ corresponding to the
ring magnet itself, given by $\rho=a$ and $z=0$,
which can not be occupied by the point magnet.
The region of space around the ring magnet, as described
by the parameter $k$ in terms of $\rho$ and $z$
is illustrated in Fig.~\ref{ksquare-fig}. Using
the leading order contributions
in Eqs.(\ref{pi3s-3t}) and (\ref{pi5s-3t}),
\begin{subequations}
\begin{eqnarray}
\pi_3(k) &=& \frac{\pi}{2} \Big[1 +{\cal O}(k^2) \Big], \\
\pi_5(k) &=& \frac{\pi}{2} \Big[1 +{\cal O}(k^2) \Big],
\end{eqnarray}
\end{subequations}
and Eq.(\ref{pi35s-3t}),
\begin{equation}
\left( \pi_5(k) -\frac{2}{k^2}
\Big\{ \pi_5(k) -\pi_3(k) \Big\} \right)
= \frac{\pi}{2} \Big[0 
%-\frac{5}{8} k^2 -\frac{35}{32} k^4
+{\cal O}(k^2) \Big],
\end{equation}
and $\lim_{\rho\to 0}k^2/\rho = 4a/(z^2+a^2)$,
in Eq.~(\ref{Eie-56}),
we reproduce the interaction energy in Eq.~(\ref{ier=034}) successfully,
for this particular case. This serves as a partial check
for the exact expression in Eq.~(\ref{Eie-56}).

For the special case when the orientation of the point
magnet is parallel to the axis of the ring magnet we have
\begin{eqnarray}
&& U(z,\rho,\phi-\phi_1,0)
= \frac{\mu_0}{4\pi} \frac{m_1(2\pi\lambda_2)}{a^2}
\left( \frac{a}{\rho} \right)^\frac{3}{2} \frac{k^3}{8}
\hspace{10mm} \nonumber \\ && \hspace{20mm} \times
\frac{2}{\pi} \left[ \pi_3(k)
-\frac{3z^2k^2}{4a\rho} \pi_5(k) \right]
\label{Eie-56-th0}
\end{eqnarray}
for arbitrary position of the point magnet.
Observe that it is independent of the variable $\phi$
representing the azimuth angle of the position of the 
point magnet leading to axial symmetry,
in addition to the trivial independence in orientation
variable $\phi_1$ because of $\theta_1=0$. Further, we have
\begin{equation}
U\left(z,\rho,\frac{\pi}{2},\theta_1\right)
=\cos\theta_1 \,U(z,\rho,\phi-\phi_1,0).
\label{Eih-60-th0}
\end{equation}
The interpretation is that,
when the azimuthal plane of position of the point dipole
is perpendicular to the azimuthal plane of its orientation,
the energy is simply a scaled version of
an axially oriented point magnet.
As a consequence of Eq.\,(\ref{Eih-60-th0}) we have the
interaction energy to be zero when the orientation of the point
magnet is perpendicular to the position vector
of the point magnet, $\theta_1=\pi/2$. That is,
\begin{equation}
U\left(z,\rho,\frac{\pi}{2},\frac{\pi}{2}\right) =0.
\end{equation}
Next, if we have $\theta_1=\pi/2$ with arbitrary 
$\phi-\phi_1$ we have 
\begin{eqnarray}
U\left(z,\rho,\phi-\phi_1,\frac{\pi}{2}\right)
= \frac{\mu_0}{4\pi} \frac{m_1(2\pi\lambda_2)}{a^2}
\sqrt{\frac{a}{\rho}} \frac{3zk^5}{32\rho} \cos(\phi-\phi_1)
\nonumber \\ \times \frac{2}{\pi} 
\Bigg[ \pi_5(k) +\frac{a}{\rho} \left( \pi_5(k) -\frac{2}{k^2}
\Big\{ \pi_5(k) -\pi_3(k) \Big\} \right) \Bigg]. \hspace{8mm}
\label{Eie-56-phph10}
\end{eqnarray}

%-----------------------------------------------------
\section{Conclusion and outlook}

In Eq.~(\ref{Eie-56}) we have presented an exact expression
for the magnetostatic interaction energy between a point magnet
and a ring magnet in terms of complete elliptic integrals.
Starting from this energy expression we can
analyze the stability of the point magnet.
Our configuration is essentially that of 
a massless point-like Levitron$^\text{TM}$,
the stability analysis of which
has been discussed in Ref.~\cite{Berry:1996sta}.
However, the investigation in Ref.~\cite{Berry:1996sta}
is assumed to be on the axis of symmetry. Our expression for
energy derived here allows an accurate analytical derivation
of the stability. This requires us to find the force on the
point dipole, which is given in terms of the derivatives of the
elliptic integrals in the energy. However, to find the
stability points this would amount to finding the zeros
of an expression involving elliptic integrals. This will
inevitably force us to depend on numerics. However,
since the stability points are expected to be close to the
axis we will be able to depend on the series expansions
and obtain analytic perturbative expressions.
This will be explored in another discussion elsewhere.

Our primary long-term goal is to discuss Casimir levitation,
as proposed in and around FIG.~16 of Ref.~\cite{Marchetta:2020dap}.
Here we outline how the methodology presented here can
be immediately used to derive the corresponding Casimir-Polder
interaction energy between a polarizable atom of polarizability
\begin{equation}
{\bm\alpha} = \alpha_1 \hat{\bf n} \hat{\bf n}
\end{equation}
and
a polarizable ring of radius $a$ with electric susceptibility
\begin{equation}
{\bm\chi} = \sigma_2 \hat{\bf z} \hat{\bf z}
\delta(z^\prime-0) \delta(\rho^\prime-a).
\end{equation}
Here $\hat{\bf n}$ is the principal axis of polarization and is
chosen to be given using Eq.\,(\ref{def-nuv}).
Similarly, $\hat{\bf z}$ is the direction of polarization of the
ring. The position of the atom is ${\bf r}$ and chosen to be
given using Eq.\,(\ref{def-rpv}), and a point on the ring is
described by ${\bf r}^\prime$ given using Eq.\,(\ref{def-rppv}),
Thus, the parameters in the problem are equivalent to those of
the magnetic configuration presented in this article.
The Casimir-Polder interaction energy between the atom and the
ring is given using Eq.\,(41) in Ref.~\cite{Marchetta:2020dap},
which can rewritten in terms of the parameters in this article as
\begin{eqnarray}
U &=& -\frac{\hbar c}{32\pi^2} \int d^3x
\left[
13 \frac{\text{tr} ({\bm\alpha} \cdot {\bm\chi})}{R^7}
-56 \frac{({\bf R} \cdot {\bm\alpha} \cdot {\bm\chi} \cdot{\bf R})}{R^9}
\right. \nonumber  \\ && \hspace{22mm} \left.
+63 \frac{({\bf R} \cdot {\bm\alpha} \cdot {\bf R})
({\bf R} \cdot {\bm\chi} \cdot {\bf R})
}{R^{11}} \right],
\end{eqnarray}
where the vector ${\bf R}$ is given by Eq.\,(\ref{def-Rrrp})
and the magnitude $R$ is given by Eq.\,(\ref{def-Rmag}).
In Ref.~\cite{Marchetta:2020dap} the atom was confined on the 
symmetry axis and it led to the significantly simplified 
expression for energy in Eq.\,(103) there. When we do not
restrict the atom to be on the axis of symmetry we have
the expression for energy
\begin{eqnarray}
U(z,\rho,\phi-\phi_1,\theta_1) = -\frac{\hbar c \,\alpha_1\sigma_2a}{32\pi^2}
\int_0^{2\pi} d\phi^\prime \left[
13 \frac{ (\hat{\bf n} \cdot \hat{\bf z})^2}{R^7}
\right. \nonumber \hspace{1mm}  \\ \left.
-56 \frac{({\bf R} \cdot \hat{\bf n}) (\hat{\bf n} \cdot \hat{\bf z})
(\hat{\bf z} \cdot {\bf R})}{R^9}
%\right. \nonumber \hspace{35mm} \\ \left.
+63 \frac{({\bf R} \cdot \hat{\bf n})^2
(\hat{\bf z} \cdot {\bf R})^2}{R^{11}} \right], \hspace{3mm} ~
\label{cas-intE}
\end{eqnarray}
where $(\hat{\bf n} \cdot \hat{\bf z})$,
$({\bf R} \cdot \hat{\bf z})$, and
$({\bf R} \cdot \hat{\bf n})$,
are given using Eqs.\,(\ref{ndz=ct1}),
(\ref{Rdz=z}), and (\ref{ndR=ex}), respectively.
The expression for energy in Eq.\,(\ref{cas-intE}) is the 
analog of our expression for magnetostatic energy in Eq.\,(\ref{ie-ex34}).
Using the methods used in this article we believe that the
three integrals in $\phi^\prime$ can be completed in terms
of elliptic integrals. The results will be reported in
a separate discussion elsewhere. 

%--------------------------------------------

\acknowledgments
A major part of this calculation was carried out during weekly
virtual meetings held on Zoom in 2021 Summer. 
We thank regular participants and the occasional visitors
for comments and collaboration. We thank Venkat Abhignan,
Anurag Kurumbail, Summer Harris, Natalie Cavallo, Ram Narayanan,
Matthew Gorban, Dylan Kelly, and Zeid Ghalyoun,
for valuable feedback.

%-----------------------------------------------------
\bibliography{biblio/b20131003-casimir-top,%
biblio/b20210330-levi}
%\nocite{*} %%% Will print the complete bib data.
%-----------------------------------------------------

\end{document}